\newif\ifsubmode
\submodefalse

\newif\ifprintfig
\printfigtrue

\ifsubmode
  \documentstyle[12pt,aasms4,epsf]{article}
  \received{}
  \accepted{}
  \journalid{}{}
  \articleid{}{}
\else
  \documentstyle[11pt,aaspp4,epsf]{article}
\fi

\lefthead{Shapiro et al.}
\righthead{}

\newcommand{\etal}{{et al.~}}

\newcommand{\kms}{\>{\rm km}\,{\rm s}^{-1}}

\newcommand{\pc}{\>{\rm pc}}
\newcommand{\Msun}{\>{\rm M_{\odot}}}

\begin{document}

\title{Observational Constraints on Disk Heating as a 
Function of Hubble Type}

\author{Kristen L. Shapiro\footnote{Also summer student and visitor 
at Space Telescope Science Institute}}
\affil{Department of Astronomy, Williams College, 
	Williamstown, MA 01267, USA}

\author{Joris Gerssen, Roeland P. van der Marel}
\affil{Space Telescope Science Institute, 3700 San Martin Drive,
       Baltimore, MD 21218, USA}



\ifsubmode\else
\clearpage\fi


\ifsubmode\else
\baselineskip=14pt
\fi


\begin{abstract}

Current understanding of the secular evolution of galactic disks
suggests that this process is dominated by two or more ``heating''
mechanisms, which increase the random motions of stars in the disk.
In particular, the gravitational influence of giant molecular clouds
and irregularities in the spiral potential have been proposed to
explain the observed velocity dispersions in the solar
neighborhood. Each of these mechanisms acts on different components of
the stellar velocities, which affects the ratio $\sigma_z/\sigma_R$ of
the vertical and radial components of the stellar velocity dispersion
since the relative strengths of giant molecular clouds and spiral
irregularities vary with Hubble type.  A study of $\sigma_z/\sigma_R$
as function of Hubble type has the potential to provide strong
constraints on disk heating mechanisms.  We present major and minor
axis stellar kinematics for four spiral galaxies of Hubble type from
Sa to Sbc, and use the data to infer the ratios $\sigma_z/\sigma_R$ in
the galaxy disks. We combine the results with those for two galaxies
studied previously with the same technique, with Milky Way data, and
with estimates obtained using photometric techniques. The results show
that $\sigma_z/\sigma_R$ is generally in the range 0.5--0.8. There is
a marginally significant trend of decreasing $\sigma_z/\sigma_R$ with
advancing Hubble type, consistent with the predictions of disk heating
theories. However, the errors on individual measurements are large,
and the absence of any trend is consistent with the data at the
1-$\sigma$ level. As a byproduct of our study, we find that three of
the four galaxies in our sample have a central drop in their stellar
line-of-sight velocity dispersion, a phenomenon that is increasingly
observed in spiral galaxies.

\end{abstract}


 
\clearpage


\section{Introduction}
\label{s:intro}

As cosmological simulations of galaxy formation are becoming
increasingly powerful, it has become apparent that these models face
their most stringent tests when compared with disk galaxies (e.g.,
Navarro \& Steinmetz 2000). While data vs. N-body
model comparisons are currently still restricted to global properties
and scaling laws (e.g., Tully-Fisher), it will soon become possible to
make more detailed comparisons.  In order to constrain these models,
it is necessary to have a detailed understanding of the internal
dynamics of disk galaxies.

Observationally, the dynamical history of a galactic disk can be
probed through the velocity distribution of its constituent stars.  In
the solar neighborhood, where the three-dimensional distribution of
stellar motions can be observed directly, it has long been established
that the stellar velocity dispersions correlate with stellar age
(i.e., spectral type).  Here, the dispersions are observed to vary from
about $\sim 10 \kms$ for the earliest spectral type stars to several
tens of $\kms$ for the latest type stars (e.g., Wielen 1977; Dehnen \&
Binney 1998).  This increase in random motion over the lifetime of a
star is often referred to as ``disk heating'' and is attributed to
scattering processes.  Spitzer \& Schwarzschild (1951) identified
giant molecular clouds as a scattering source while Barbanis \&
Woltjer (1967) recognized that stars are also perturbed by
irregularities in the potential associated with spiral arms.  The
former of these mechanisms increases the velocity dispersions
isotropically, while the latter primarily increases the dispersions in
the plane of the disk.  Consequently, different combinations of spiral
structure and molecular clouds lead to different relative amounts of
heating in the radial and vertical directions of a galactic disk.
This was demonstrated explicitly by Jenkins \& Binney (1990)
using numerical simulations.

Disk heating can be quantified by what is known as the velocity
ellipsoid.  Schwarzschild (1907) showed that the stellar velocities in
the solar neighborhood are well described by a trivariate Gaussian
distribution.  Since the velocity dispersions near the sun are
observed to obey $\sigma_R > \sigma_\phi > \sigma_z$, Schwarzschild's
distribution has an ellipsoidal shape in velocity space.  The exact
shape of this ellipsoid is defined by its two axis ratios,
$\sigma_z/\sigma_R$ and $\sigma_\phi/\sigma_R$.  (The orientation of
the velocity ellipsoid is set by the cross-terms $\sigma_{Rz}$ and
$\sigma_{R\phi}$, which, in the plane of the disk, are both zero for a
stellar population that is distributed symmetrically.)  In an
axisymmetric disk with stellar orbits not too far from circular (the
epicycle approximation), $\sigma_\phi/\sigma_R$ depends only on the
circular velocity and not on any disk heating mechanism. Measurements
of the ratio $\sigma_z/\sigma_R$ therefore constrain the heating
processes in galactic disks.

The dependence of disk heating on the relative amount of molecular gas
and spiral structure suggests that the velocity ellipsoid ratio
$\sigma_z/\sigma_R$ should vary as a function of Hubble type, since
both molecular gas content and morphology are functions of Hubble
type.  This hypothesis was first tested by Gerssen et al.~(1997,
2000), who developed a method for measuring all three components of
the stellar velocity dispersions in external galaxies.  Their
application of this technique to NGC~488 (Sb) and NGC~2985 (Sab)
showed that the derived ratios in these systems are consistent with
the trend seen in simulations by Jenkins \& Binney (1990).  The same
is true for the Milky Way (Sbc).  However, for the external galaxies,
the errors on each individually-determined velocity ellipsoid ratio
are rather large, since galactic disks are characterized by low
surface brightness and by relatively small internal velocity
dispersions.  Rather than try to reduce the errors on individual
points, we attempt here to improve the statistics on a possible relation
with Hubble type by applying the technique of Gerssen et al. to four
additional disk galaxies.  The inclusion of these data both triples
the sample of galaxies with direct measurements of the stellar
velocity ellipsoid ratio and expands the sampled range of
morphological types.

\section{Method}
\label{s:theory}

In an intermediate-inclination galaxy, spectra obtained along radius
vectors with different position angles will probe different projections
of the velocity ellipsoid.  The velocity dispersions along an
arbitrarily positioned spectrum in a thin axisymmetric disk can be
expressed as
\begin{equation}
\sigma_{\rm los}^2 = \left[ \sigma_R^2 \sin^2 \theta + \sigma_\phi^2 \cos^2 
  \theta \right] \sin^2 i + \sigma_z^2 \cos^2 i ,
\label{e:los}
\end{equation}
where $i$ is the inclination angle of the galaxy and $\theta$ is the
position angle with respect to the major axis in the plane of the
disk.  In this equation, there are three unknown quantities
($\sigma_R$, $\sigma_\phi$, and $\sigma_z$), but $\sigma_\phi/\sigma_R$
is constrained by the epicycle approximation (see below).  Hence,
there are two independent unknowns in this equation, and observations
along at least two axes are required to extract both (Gerssen et
al. 1997).  Maximum leverage is obtained by taking spectra along the
major axis, where the line-of-sight stellar velocity dispersions are a
combination of $\sigma_\phi$ and $\sigma_z$, and along the minor axis,
where they are a combination $\sigma_R$ and $\sigma_z$.  In an
axisymmetric disk, the stellar velocities along any third axis will
simply be a linear combination of these two and will not provide
additional information.  

Assuming that most orbits in disk galaxies are well-described by the
epicycle approximation, the dispersions in the azimuthal and radial
direction are related according to
\begin{equation}
\frac {\sigma_\phi^2}{\sigma_R^2} = \frac{1}{2} \left( 1 +
  \frac {\partial \ln V_c}{\partial \ln R} \right),
\label{e:gamma}
\end{equation}
where $V_c$ is the circular speed in the galaxy.  This equation,
together with the observations along the major and minor axes, fully
constrains the velocity ellipsoid ratio $\sigma_z/\sigma_R$.  This
model is illustrated in Figure~\ref{f:figsigrat}, in which the
predicted major to minor axis velocity dispersion ratio $\sigma_{\rm
major}/\sigma_{\rm minor}$ is plotted against the velocity ellipsoid
ratio $\sigma_z/\sigma_R$ for a variety of inclinations under the
assumption of a flat rotation curve ($V_c$ = constant).  In principle,
any value of $\sigma_z/\sigma_R$ can be obtained using this method,
but disk heating theories as well as solar neighborhood observations
suggest that this ratio, in practice, does not exceed unity.

When direct measurements of the circular speed are available
(e.g. from an emission-line rotation curve), this information can be
included in equation~(\ref{e:gamma}).  Otherwise, $V_c$ must be
estimated from the stellar rotation curve $\overline{V}(R)$ through
the asymmetric drift equation,
\begin{equation}
V_c^2 - \overline{V}^2 = -R \frac{\partial \sigma_{Rz}}{\partial z} +
\sigma_R^2 \left[\frac{R}{h} - R \frac{\partial}{\partial R}
\ln(\sigma_R^2) - \frac{1}{2} + \frac{R}{2V_c} \frac{\partial
V_c}{\partial R} \right] ,
\label{e:asymdrift}
\end{equation}
where $h$ is the photometric disk scale length and $\sigma_{Rz}$
measures the tilt of the velocity ellipsoid in the ($R$,$z$) plane.
The tilt term $R\frac{\partial \sigma_{Rz}}{\partial z}$ is always
between 0 and ($\sigma_R^2 - \sigma_z^2$) (Binney \& Tremaine 1987).
(Orbit integration by Binney \& Spergel (1983) and Kuijken \& Gilmore
(1989), see also Kent \& de Zeeuw (1991), suggests that the solar
neighborhood value is midway between these two extremes.)
Substitution of these two limiting cases of the tilt term into the
model does not produce significantly different results (Gerssen \etal
1997).  For the present study, this term was set to zero without loss
of generality.  When both gas and stellar rotation curves are
available, the system is overdetermined, and the asymmetric drift
equation can function either as a consistency check on the derived
model velocities or as an additional constraint to more consistently
fit all the available data.

Although the functional forms of all the variables can be directly
determined using the above equations, this method is very sensitive to
noise (Merrifield \& Kuijken 1994).  Instead, we adopt a model-fitting
approach.  We assume that, at the radii where the disk light
dominates, the circular velocity can be described by a power law,
\begin{equation}
V_c = V_{c,0} R^\alpha .
\end{equation}
The other two observables, $\sigma_{\rm major}$ and $\sigma_{\rm
minor}$, are modeled assuming exponential distributions for both the
radial and the vertical dispersion components.
\begin{equation}
\sigma_R = \sigma_{R,0} \ e^{-R/h_{\rm kin}} ,
\end{equation}
\begin{equation}
\sigma_z = \sigma_{z,0} \ e^{-R/h_{\rm kin}} .
\end{equation}
Although there is no a priori reason to assume that the kinematical
scale length, $h_{\rm kin}$, should be the same for both the radial and
the vertical component, the quality of our data is insufficient to
constrain both of these parameters independently.

\section{Observations and Data Reduction}
\label{s:obs}

\subsection{Spectroscopy}

Observations were carried out on January 12--15, 2002 at the Mayall
4-meter telescope at the Kitt Peak National Observatory.  We observed
four galaxies: NGC 1068, NGC 2460, NGC 2775, and NGC 4030.  All four
are nearby, spiral galaxies of intermediate inclination ranging in
Hubble type from Sa to Sbc.  The main characteristics of each galaxy
are listed in Table~\ref{t:sample}.

Long-slit spectra were obtained along the photometric major and minor
axes of each galaxy, in multiple exposures of 30 minutes each.  The
orientation and position of each slit is shown schematically in
Figure~\ref{f:figsbdata}, and the log of the observations is reported
in Table \ref{t:obslog}.  Spectra of several late-G and early-K giant
stars were also acquired to serve as templates in the stellar
kinematical analysis. All data were taken with the Ritchey-Chretien
Focus Spectrograph using the KPC-24 grating in second order, centered
on the Mg b absorption feature at $\sim 5175$~\AA, which yielded a
velocity scale of $30~\kms$ per pixel on the T2KB CCD.  The spatial
scale of this configuration is 0.69 arcsec/pixel.  The long-slit
spectra (5.4 arcmin long) were obtained with a 3.0 arcsec wide slit.
Seeing conditions varied during the observations but were fairly poor,
$\sim 1.5$ arcsec.  However, our kinematical analysis of the data
focuses on smooth kinematical gradients at large radii, and poor
spatial resolution is therefore not a problem.  Calibration arc
spectra obtained with a FeAr lamp were taken every 30 minutes.

The data were reduced using the standard IRAF\footnote{IRAF is the
Image Analysis and Reduction Facility distributed by the National
Optical Astronomy Observatories.}  packages.  Each exposure was
overscan- and bias-subtracted, corrected for pixel-to-pixel variations
using internal lamp exposures, and corrected for large scale gradients
and vignetting using sky frames obtained in evening or morning
twilight.  The arc lamp frames were used to wavelength-calibrate and
spatially rectify the spectra.  All spectra were rebinned onto a
logarithmic wavelength scale to facilitate the kinematical analysis.
Finally, all spectra were sky subtracted using data near the edges of
the slit.

In each galaxy spectrum, rows along the spatial direction were
co-added to increase the signal-to-noise ratio.  These co-added bins
ranged in size from one pixel in the center of a galaxy to $\sim 20$
pixels at large radii.  For the stellar template spectra, the relevant
rows along the slit were combined to create one-dimensional stellar
spectral profiles.

\subsection{Photometry}
\label{s:photo}

In addition to the spectroscopic data, we obtained photometric data
from several archives.  For NGC~2460 and NGC~4030, {\it HST}/WFPC2
images in filter F814W were obtained from the {\it HST} archive.
Archived {\it K} band images were obtained for NGC~1068 (2MASS) and
NGC~2775 (Mollenhoff \& Heidt 2001).  After sky subtraction, the
azimuthally averaged surface brightness profiles were extracted for
each galaxy using the stsdas task ELLIPSE in IRAF. The results are
illustrated in Figure~\ref{f:figsbdata}.

At large radii, all four of the galaxies exhibit exponential disks,
the scale lengths of which are found by simply fitting a line to this
part of the surface brightness profile.  Furthermore, subtraction of
the linear fit from the profile shows the radii at which light from
the bulge dominates that of the galaxy.  At these radii, the derived
stellar velocities are a measure of the bulge kinematics and not of
the disk kinematics.  Kinematical data points obtained at these radii
were therefore excluded from the modeling.  Estimates of the
inclination of each galaxy were also found in the isophotal fitting
procedure. The derived quantities are reported and compared to
literature values in Table~\ref{t:sample}.

\section{Kinematical Analysis}
\subsection{Absorption Lines}

Stellar velocity dispersions were extracted from the reduced spectra
using the standard assumption that the observed galaxy spectrum is the
convolution of a Gaussian velocity distribution and a typical stellar
(template) spectrum.  We did not attempt to extract higher-order
moments of the velocity distributions after a preliminary analysis
revealed no significant departures from a single Gaussian.

The Gaussian that best fits the galaxy data in a least-squares sense
yields the mean stellar line-of-sight velocities and the stellar
velocity dispersions.  We derived the stellar kinematics using two
independent methods, one of which performs the least square
minimization in pixel space (van~der~Marel 1994) and the other of
which does this task in Fourier space (using software originally
developed by Konrad Kuijken). The results derived from both methods
were found to be in good agreement.  The analysis was repeated with
each template star.  Different template stars did not yield
significantly different results and we therefore did not attempt to
construct an optimal template spectrum. Instead, we used the template
star that yielded the the marginally best $\chi^2$ value.  Three of
the four galaxies (NGC~1068, NGC~2460, and NGC~4030) were best modeled
by HD55184, of spectral type K0III; the fourth, NGC~2775, was most
accurately fit by HD18991, of type G9III.  The derived major and minor
axis mean stellar velocity and velocity dispersions for each galaxy
are shown in Figure~\ref{f:figkin}.  The abscissa is the radius in the
disk, which for the minor axis was obtained through deprojection.

Although the primary focus of this project is on the outer disk
kinematics, we note that three of the four galaxies in our sample show
noticeable central drops in their stellar velocity dispersions.  Such
features are increasingly observed in kinematical studies of both
Seyfert and non-active spiral galaxies (Bottema 1989; Emsellem \etal
2001; de~Zeeuw \etal 2002; Marquez \etal 2003).  This phenomenon is
generally attributed to a kinematically decoupled core (Bottema \&
Gerritsen 1997).  Alternatively, Emsellem \etal (2001) suggested that
these central dispersion drops are the results of recent star
formation in nuclear disks.  They note that such a process could be
fueled by gas inflow along a nuclear bar.  However, the three galaxies
in which we detect this phenomenon (NGC~2460, NGC~2775, and NGC~4030)
are non-active and are not known to display starburst activity
(although such activity may have occurred in the recent past).  Other
explanations may therefore be necessary to account for the central
dispersion drops in these galaxies.

\subsection{Emission Lines}
\label{s:emission}

The spectral range of the long-slit spectra also included several
emission lines, most notably the [OIII] line at 5007~\AA.  For three
of our four galaxies, these [OIII] lines were fitted with a Gaussian
profile, using the same spatially co-added bins as for the absorption
line data.  Due to the Seyfert nature of the fourth galaxy, NGC~1068,
the [OIII] line is completely saturated in our spectra. Instead, for
this galaxy we used the [NI] doublet at 5199~\AA.  The two emission
lines in this feature were each fitted simultaneously with a Gaussian
profile and the combined fit accurately matched both of the emission
lines.  The wavelength shift between the two Gaussians was kept fixed
since we assume that both lines trace the same kinematics.

Since the velocity dispersions of gas in disks is typically $\lesssim
10 \kms$, the rotation curves derived in this manner provide a direct
estimate of the circular velocities of the galaxies.  The major axis
gas rotation curves are overplotted on the stellar rotation curves in
Figure~\ref{f:figkin}.  Except for NGC~4030, the asymmetric drift
(i.e.  the difference between the gas and the stellar velocities) is
clearly noticeable.  Apparently, the gas in NGC~4030 is not an
accurate tracer of the circular velocity in this system.  Also unusual
is the central feature in the gas kinematics of NGC~1068, which is
most likely due to the bar and complicated structure in the nucleus of
this system (Scoville \etal 1988).  At larger radii, however, the gas
kinematics for this galaxy are quite regular.

\section{Modeling}
\label{s:modeling}

We used the model described in Section \ref{s:theory} to
simultaneously fit the three observables, $\sigma_{\rm major}$,
$\sigma_{\rm minor}$, and $\overline{V}(R)$.  The five best-fit model
parameters, $\sigma_{R,0}$, $\sigma_{z,0}$, $h_{\rm kin}$, $V_{c,0}$,
and $\alpha$ were obtained using the non-linear fit routines of Press
\etal (1992).  The inferred parameters for each galaxy are tabulated
in Table~\ref{t:allresults}, and the corresponding best model fits are
shown in Figure~\ref{f:figallmodels} over the range of radii for which
the kinematics are dominated by the disk.  All results were confirmed
using an alternative minimization routine, the downhill simplex method
(Press \etal 1992).  Error bars on the best-fitting model parameters
were determined using bootstrapping.

We examined the dependence of the derived parameters, and in
particular the velocity ellipsoid ratio $\sigma_z/\sigma_R$ on
uncertainties in the input parameters.  Within the error ranges on the
photometric scale length $h$ and the inclination $i$ quoted in
Table~\ref{t:sample}, changes in the values of these two parameters do
not significantly affect the final results.  We also tested the
dependence on position angle by assuming that the true major axis is
offset from our observed major axis by $\Delta {\rm PA}$ degrees.  We
then reran the models and found that mismatches of up to 25 degree in
the adopted major axis position angle do not significantly affect the
velocity ellipsoid ratio at a level beyond $\sim 0.1$.  As is evident
from the major and minor axis rotation curves in
Figure~\ref{f:figkin}, none of the galaxies in our sample have major
axis position angles that were misidentified by more than $\sim 10$
degrees.

Observations of edge-on galaxies show that the scale height of a disk
is generally independent of distance from the galaxy center and that
the vertical surface brightness distribution can be fitted with a
hyperbolic secant function (e.g. van~der~Kruit 1989).  The isothermal
sheet approximation, together with these observations implies that the
vertical component of the stellar velocity dispersion $\sigma_z$
should decline exponentially with radius with a kinematical scale
length $h_{\rm kin}$ that is twice the photometric scale length $h$.
However, in our model fits we do not generally find kinematical scale
lengths that obey this relation.  So most likely the assumptions on
which $h_{\rm kin}=2h$ is based do not hold in these galaxies (or in
general).

In the following sections, details of the modeling procedure for each
galaxy are described.

\subsection{NGC 1068}

NGC~1068 is one of the prototypical Seyfert galaxies.  Its central
structure is quite complex. High spatial resolution CO observations
(Schinnerer \etal 2000) show both a central bar and central spiral
arms. The central feature in our emission line rotation curve
(Figure~\ref{f:figkin}, panel a) is another manifestation of the
central complexity in this system.  However, at larger radii the
morphology is similar to that of a regular Sb galaxy.  As the
brightest and nearest of the four galaxies in our sample, NGC~1068 has
the most extensive stellar kinematical data set of the four galaxies.
Its kinematics are accurately fit by the model.  In contrast, the gas
kinematics are more erratic and are measured to smaller radial extent,
but they are not inconsistent with the circular velocities predicted
from the stellar rotation curve using equation~(\ref{e:asymdrift})
(dotted line in Figure~\ref{f:figallmodels}).  Kaneko \etal (1992),
using several optical emission lines, derive a gas rotation curve that
corresponds closely to the model prediction of $V_{c,0}$.  The
best-fit model has velocity ellipsoid axis ratio $\sigma_z/\sigma_R$ =
$0.58 \pm 0.07$.

As shown in Figure~\ref{f:figsbdata}, the large radial extent of NGC
1068 dictates that the diameter of the galaxy along the major axis is
somewhat longer than the slit length.  However, near the slit ends,
the night sky completely dominates the underlying galaxy light, and
this part of the slit is therefore very effective in subtracting the
sky background.  Indeed, our kinematical spectral modeling provided no
evidence of residual night sky contamination.

\subsection{NGC 2460}
\label{s:2460mod}

Panel (b) in Figure~\ref{f:figkin} shows that the amplitudes of the
stellar rotation curves on the two opposite sides of NGC~2460 are
somewhat different.  This disparity is probably related to a
significant asymmetry in the galaxy's HI distribution (Haynes
\etal 1998).  As the only Sa type galaxy in our sample, NGC 2460 has a
relatively large bulge, which means that much of our kinematical data
were discarded on the grounds that they are dominated by the bulge
kinematics.  This left few disk data points, especially along the
minor axis.  To compensate for these limitations, we tried including
the gas kinematics in the model fit as a constraint on the circular
velocity (instead of predicting $V_c(R)$ from $\overline{V}(R)$ using
equation~\ref{e:asymdrift}).  We also tried separate fits using data
sets that included only one or both sides of the stellar rotation
curve.  There is considerable scatter in the individual fits, and the
velocity ellipsoid ratio $\sigma_z/\sigma_R$ is therefore not very
precisely determined for NGC~2460.  We find $\sigma_z/\sigma_R$ =
$0.83 \pm 0.35$, with the large error bar representing the scatter
among the results of the different ways of performing the data-model
comparison.  Observations of the rotation curve of NGC~2460 by Marquez
\etal (2002), when deprojected, yield a circular speed of
approximately 210 $\kms$.  This compares well to our observations,
which suggest a similar value, and to our model fit of 218 $\pm$ 24
$\kms$.

\subsection{NGC 2775}

Most often classified as either Sa or Sab, NGC~2775 has a large bulge
whose exact extent is a matter of some debate in the literature (see
below).  To best fit the kinematics of this galaxy, we assumed a flat
rotation curve, as has been observed to large radii by Rubin \etal
(1985) and Corsini \etal (1999).  Figure~\ref{f:figallmodels} shows
the result obtained under this assumption.  The models are able to fit
most data except for the major axis dispersions at the largest radii
(which remain rather flat and therefore try to push the kinematical
scale length to infinity).  The derived value of the velocity
ellipsoid axis ratio is $\sigma_z/\sigma_R$ = $1.02 \pm 0.11$.

Rubin \etal (1985) and Corsini \etal (1999) both find the circular
speed to be $\sim$310 $\kms$, a value that is somewhat higher than our
measured gas velocity of $\sim$280 $\kms$.  As our circular speed was
derived using a fairly weak [OIII] line (while the literature values
were obtained with higher signal-to-noise), the modeling was rerun
with the circular speed forced to coincide with the literature
quantities.  This change did not alter the results significantly.

Comparison of the photometric analysis described in Section
\ref{s:photo} with the literature revealed that the bulge-disk
transition radius of NGC~2775 is a value of much debate (see Boroson
1981; Grosbol 1985; Moriondo \etal~1998; Mollenhoff \& Heidt 2001).
The two most recent decompositions, done using $K$ band data by
Moriondo \etal and by Mollenhoff \& Heidt, disagree significantly.
The decomposition of Mollenhoff \& Heidt concurs with those of Boroson
(1981) and Grosbol (1985) in concluding that the disk dominates the
galactic light only outwards of $\sim$70 arcsec.  In contrast,
Moriondo et al. quote an effective bulge radius of only 14 arcsec.
This last result resembles that found in this project, which restricts
the bulge's influence to the inner 30-40 arcsec of the galaxy.  Given
the disparity in the literature, the possibility exists that the bulge
dominates the dispersions even at radii where we assumed them to be
disk dominated.  Since bulges of galaxies are believed to be roughly
isotropic (Kormendy \& Illingworth 1982; Spaenhauer \etal 1992), the
inferred velocity ellipsoid ratio $\sigma_z/\sigma_R$ must therefore
be considered an upper limit.

\subsection{NGC 4030}

A complication in the modeling of NGC~4030 is the gas rotational
velocity, which is similar to that of the stars and which therefore
indicates that the gas cannot be rotating at the circular velocity.
(The alternative explanation that the stellar rotation in NGC~4030 is
close to the circular velocity is unlikely since the observed stellar
velocity dispersions in this system are comparable in magnitude to the
dispersions observed in the other galaxies in the sample.)  This may
result from the gas not being in an equilibrium state at present or
from the gas not being characterized by circular orbits, either of
which could be the result of a recent interaction.  Either way, the
gas kinematics of this galaxy can clearly not be included to help
constrain our models.  We therefore fitted the model to the stellar
data to obtain $\sigma_z/\sigma_R$ = 0.64 $\pm$ 0.28.  The stellar
kinematics are well fitted by the model, but the predictions for the
circular velocity are larger than the observed gas kinematics, as
expected.  The large error on the velocity ellipsoid ratio reflects
the relatively low data quality for the minor axis data, the
observations of which were affected by reduced transmission due to
clouds.  The total integration time along this axis is effectively
only half as much as for the major axis observations.

The only literature measurement of the gas kinematics in this galaxy
is that of Mathewson \& Ford (1996), who measured the maximum
rotational velocity of the gas and found a value of 236 $\kms$.  This
is similar to the rotational speed of our data, which, when
deprojected, is $\sim$230 $\kms$.  Both of these values, however,
coincide with stellar velocities and so are lower than the predicted
circular speed of 263 $\kms$.

\section{Results}
\label{s:results}

Combining the velocity ellipsoid axis ratios derived here with the
other two direct measurements of this value for external galaxies
(Gerssen \etal 1997, Gerssen \etal 2000) and with the value for the
Milky Way yields the result shown in Figure~\ref{f:figfinal}.  This
graph represents the first direct investigation of the velocity
ellipsoid ratio as a function of morphological type.

In addition to these kinematical results, van der Kruit \& de Grijs
(1999) present an alternative method of constraining the ratio of the
velocity ellipsoid.  Their approach is based on photometric
observations of edge-on galaxies and therefore relies on a number of
assumptions in order to link the photometry to a kinematical quality.
The velocity ellipsoid ratios derived using this technique are
presented as averages over all the galaxies of a particular Hubble
type (where the sample ranges from Sb to Sd).  The earliest
types in their sample have average ratios that agree well with our
individually-determined ratios in the same bins. 

The strength of the trend of velocity ellipsoid ratio with Hubble type
was assessed both for the set of direct measurements of this value
(solid points in Figure~\ref{f:figfinal}) and for all available data
(solid and open points).  Linear regressions that account for both
horizontal and vertical errors were performed.  The fit to the set of
direct measurements of the velocity ellipsoid axis ratio has a slope
of $-0.11 \pm 0.08$.  Van~der~Kruit and de~Grijs observed no obvious
trend in their data (open symbols). Linear regression of the combined
results of their study with those of this paper indicate a slope of
$-0.05 \pm 0.04$.  For both regressions the goodness-of-fit is
statistically acceptable and the data marginally suggest a downward
trend in the ratio $\sigma_z/\sigma_R$ with advancing Hubble
type. However, this is significant only at the one sigma level, and
the absence of any trend is not ruled out by the available data.

\section{Discussion and Conclusions}

The fits described in Section~\ref{s:modeling} show that modeling each
of the galaxies in our sample is not without difficulties, despite the
optically regular-looking morphologies of the galaxies.  Problems
include the asymmetric rotation curve of NGC~2460 and the fact that
the gas velocities in NGC~4030 probably do not trace the circular
velocity.  To some extent this may simple reflect the real nature of
galaxies.  For example, early results from the SAURON survey of
elliptical galaxies show (de~Zeeuw \etal 2002) that a regular
appearance alone is no guarantee of regular stellar kinematics.

The results derived here are qualitatively consistent with the
implicit predictions of Jenkins and Binney (1990), who attribute
secular evolution in disk galaxies to a combination of spiral
irregularities and encounters with GMCs.  They express the relative
importance of these two mechanisms with a parameter $\beta$, and their
simulations indicate that increasing $\beta$ (increasing the
importance of spiral structure) results in smaller $\sigma_z/\sigma_R$
(Jenkins \& Binney 1990).  For the solar neighborhood, they use the
observed shape of the velocity ellipsoid and the velocity
dispersion-age relation to conclude that $\beta \sim 90$.

To estimate the relative importance of GMCs in the different galaxies
of our sample we used the FCRAO extragalactic CO survey (Young \etal
1995).  The H$_2$ surface densities (Table~\ref{t:gas}) were derived
assuming a constant CO to H$_2$ conversion factor.  The derived
surface densities indicate that the molecular content of the sample
galaxies decreases with advancing Hubble type.  (The exception to this
trend is NGC~1068, in which Young \etal observe both an exponential
distribution of gas and a CO ring; this combination makes the mean
surface density of H$_2$ difficult to estimate).  Various studies
(e.g. Wilson 1995) find that this factor increases as the metallicity
of the host galaxy decreases.  Using Wilson's expression and an
estimate of the metallicity in each galaxy obtained from the observed
correlation between metallicity and absolute magnitude (Roberts \&
Haynes 1994), we calculate that changes in the conversion factor due
to metallicity are less than 10 percent for the galaxies in our
sample, except for NGC~1068 where the difference is $\sim 20$ percent.
This is much less than the differences in H$_2$ density between the
galaxies in Table~\ref{t:gas}, and therefore does not affect the
present discussion at a significant level.

The morphological classifications of galaxies imply that the
earlier-type galaxies are more tightly wound and more regular (see
images in Figure~\ref{f:figsbdata}).  It is therefore quite likely
that the spiral potentials associated with earlier-type galaxies are
smoother than those of later-types.  Combined with the H$_2$ surface
densities in Table~\ref{t:gas} this suggests that the earlier-type
galaxies have smaller values for $\beta$ and thus larger
$\sigma_z/\sigma_R$ ratios.  The results of our study are consistent
with this prediction and, at 1-$\sigma$ confidence, confirm it.

An alternative explanation of disk heating is presented by Hanninen
and Flynn (2002), who attribute observed heating to the combined
influences of GMCs and hypothesized massive halo black holes.
Although simulations using these assumptions do account for the solar
neighborhood velocity ellipsoid, this need not extend to external
galaxies since the dark matter content in these systems is generally
not well known.  Any further analysis of this possibility awaits
observational confirmation of the existence of a significant
population of intermediate-mass black holes in the halos of galaxies
(see van~der~Marel 2003 for a review of present observational
constraints).

In order to provide more rigorous constraints on theoretical models,
additional data is necessary.  The technique described here is
straightforward and can readily be applied to a larger sample of disk
galaxies.  Despite the sizable error bars associated with individual
galaxies, an increase in sample size could unequivocally establish the
presence of the trend suggested by our data.  The results presented
here are only a first step in the observational investigation that is
critical to understanding disk heating.  Subsequent work could benefit
from the use of Integral Field Units.  IFUs that have a sufficiently
large field-of-view and enough spectral resolution to study stellar
velocity dispersions ($R \gtrsim 5000$) are ideally suited to extend
the sample of galaxies.  Kinematical data sets that cover a
substantial fraction of a disk will provide both a consistency check
on the underlying model assumptions and more precisely determined
constraints on the velocity ellipsoid.


\acknowledgements{
We thank Konrad Kuijken for his absorption line analysis program and
the referee for useful comments.  We are also very grateful to Kitt
Peak National Observatory for the use of their facilities and
their helpful staff, in particular Bill Gillespie.  KLS wishes to
thank the Space Telescope Science Institute for repeated hospitality
in hosting her as both a summer student and a visitor.}

\clearpage




\ifsubmode\else
\baselineskip=10pt
\fi


\clearpage


\ifsubmode\else
\baselineskip=14pt
\fi


\newcommand{\figcapsigrat}{
Relation of the ratio of the observed major to minor axis dispersions
to the velocity ellipsoid ratio $\sigma_z/\sigma_R$, assuming a flat
rotation curve.  Under this assumption, the major axis dispersions
cannot be smaller than $\sqrt{1/2}$ times the minor axis dispersions.
The velocity ellipsoid ratio $\sigma_z/\sigma_R$ can in principle have
any value; however, observations of this ratio in the solar
neighborhood and existing disk heating theories suggest that
$\sigma_z/\sigma_R$ does not exceed unity.  Different curves are for
inclinations between $0^\circ$ (face-on) and $90^\circ$ (edge-on) in
steps of 10 degrees.
\label{f:figsigrat}}

\newcommand{\figcapsbdata}{
Logarithmic grey scale images of the four galaxies in our sample are
shown in the top panels.  From left to right: NGC~1068 (DSS image),
NGC~2460, NGC~2775, and NGC~4030. The last three images are archival
{\it HST}/WFPC2 images obtained with the F814W filter.  In each panel,
North is at the top and East is to the left. The 5.4 arcmin long-slits
positions along which we obtained the spectra are overplotted in each
panel.  The slits are positioned close to the major and minor axes in
each galaxy (see text).  The panels on the bottom show the radial
surface brightness profiles derived from the images. (For NGC 2460 and
NGC 4030, the WFPC2 images shown here were used; for NGC 1068, a 2MASS
$K$ band image was used; and for NGC 2775, a $K$ band image acquired
by Mollenhoff \& Heidt (2001) was used.)  The profiles show the extent
of the bulge component in each galaxy.  The solid lines are
exponential fits to the outer disk component.  The best-fit
exponential scale lengths are in close agreement with literature
values (see Table \ref{t:sample}).
\label{f:figsbdata}}

\newcommand{\figcapkin}{
Observed kinematics of the four galaxies in our sample: {\bf a: top
left} NGC 1068, {\bf b: top right} NGC~2460, {\bf c: bottom left}
NGC~2775 and {\bf d: bottom right} NGC~4030.  In each panel the
kinematical information is given in $\kms$.  Filled circles represent
major and minor axis stellar data; crosses represent gas data.  Error
bars for the stellar velocities are shown but are generally smaller
than the plot symbols. The error bars for the gas velocities are shown
in Figure~\ref{f:figallmodels}.  The top panel for each galaxy shows
the gas major axis rotation curve, the stellar major axis rotation
curve, and the stellar minor axis velocities overplotted (the latter
are always near 0 $\kms$).  The galaxy systemic velocities have been
subtracted.  The lower two panels for each galaxy show the major and
minor axis stellar velocity dispersions respectively.  The abscissa is
the radius in the disk, which, for the minor axis, was obtained
through deprojection.
\label{f:figkin}}

\newcommand{\figcapallmodels}{
Results of fitting the model to each of the four galaxies.  All
kinematical quantities are in $\kms$.  Solid points represent stellar
data, and crosses represent gas kinematical data.  The best-fit models
for the stellar kinematics of each galaxy are shown as solid lines.
The dotted lines are the circular velocities $V_c(R)$ in the models.
For NGC~1068 ({\bf a: top left}), NGC~2775 ({\bf c: bottom left}), and
NGC~4030 ({\bf d: bottom right}), $V_c(R)$ was predicted based on the
stellar rotation curve $\overline{V}(R)$ using
equation~(\ref{e:asymdrift}).  For NGC~2460 ({\bf b: top right}), the
gas kinematics were used to constrain $V_c(R)$.
\label{f:figallmodels}}

\newcommand{\figcapfinal}{
Velocity ellipsoid ratio $\sigma_z/\sigma_R$ as a function of galactic
type (Hubble stage T).  The solid points displayed here represent the
results obtained kinematically for individual galaxies and are best
fit by the solid line.  The open points represent the photometric
results of van der Kruit \& de Grijs (1999); when combined with our
results, the data are best fit by the dashed line.  Vertical error
bars for our data are derived in this paper, and those for van der
Kruit \& de Grijs's data are statistical (the RMS of determinations
for multiple galaxies of the same type).  Horizontal errors represent
the uncertainty inherent in galaxy classification (Naim \etal 1995).
The $\sigma_z/\sigma_R$ value for NGC~2775 is an upper limit (arrow)
and was not included in the linear regression analysis.
\label{f:figfinal}}


\ifsubmode
\figcaption{\figcapsigrat}
\figcaption{\figcapsbdata}
\figcaption{\figcapkin}
\figcaption{\figcapallmodels}
\figcaption{\figcapfinal}
\clearpage
\else\printfigtrue\fi

\ifprintfig


\clearpage
\begin{figure}
\epsfxsize=15.0truecm
\epsfysize=15.0truecm
\centerline{\epsfbox{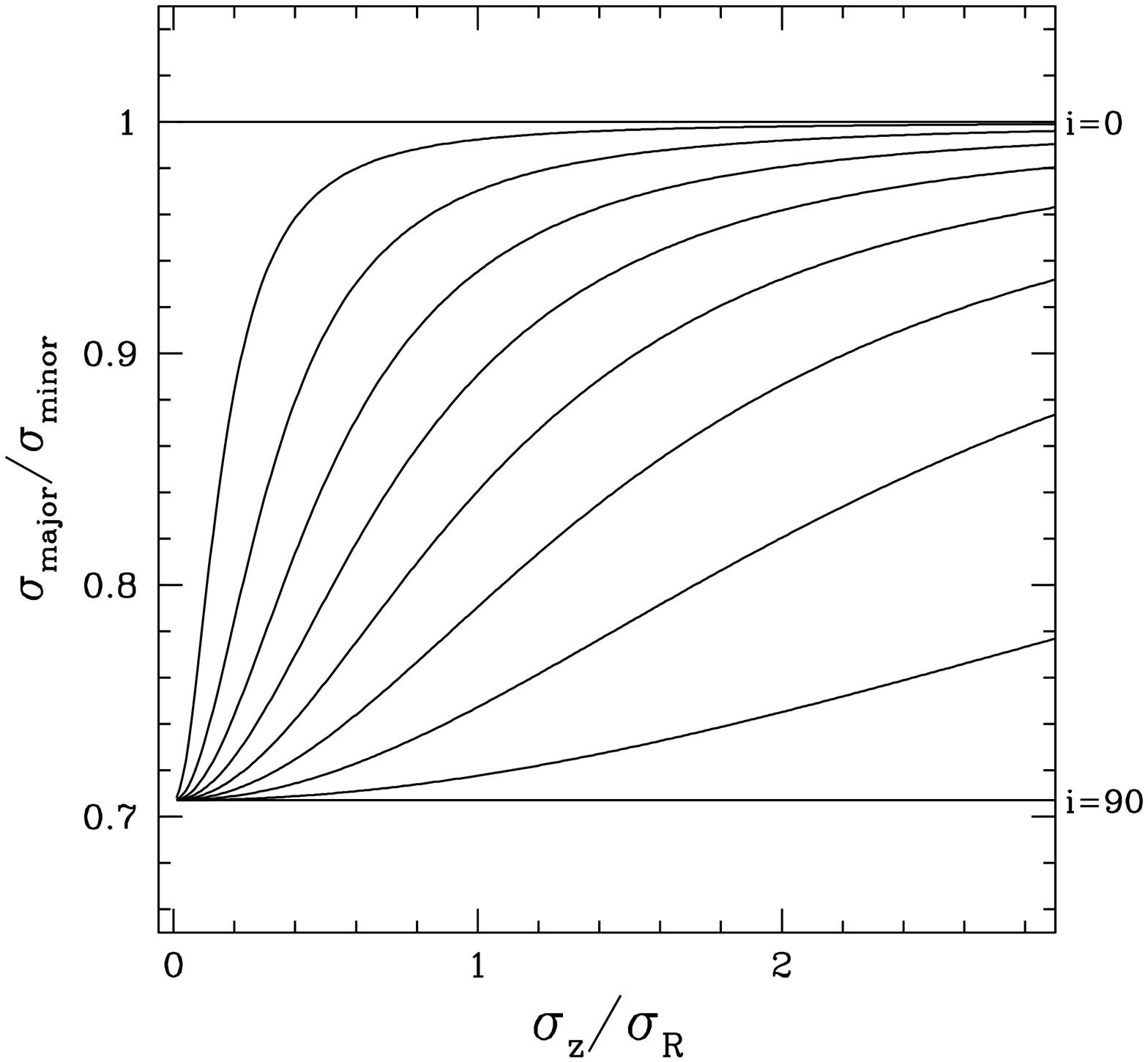}}
\ifsubmode
\vskip3.0truecm
\setcounter{figure}{0}
\addtocounter{figure}{1}
\centerline{Figure~\thefigure}
\else\figcaption{\figcapsigrat}\fi
\end{figure}

\clearpage
\begin{figure}
\epsfxsize=0.5\hsize
\hbox to \hsize{\noindent\epsfbox{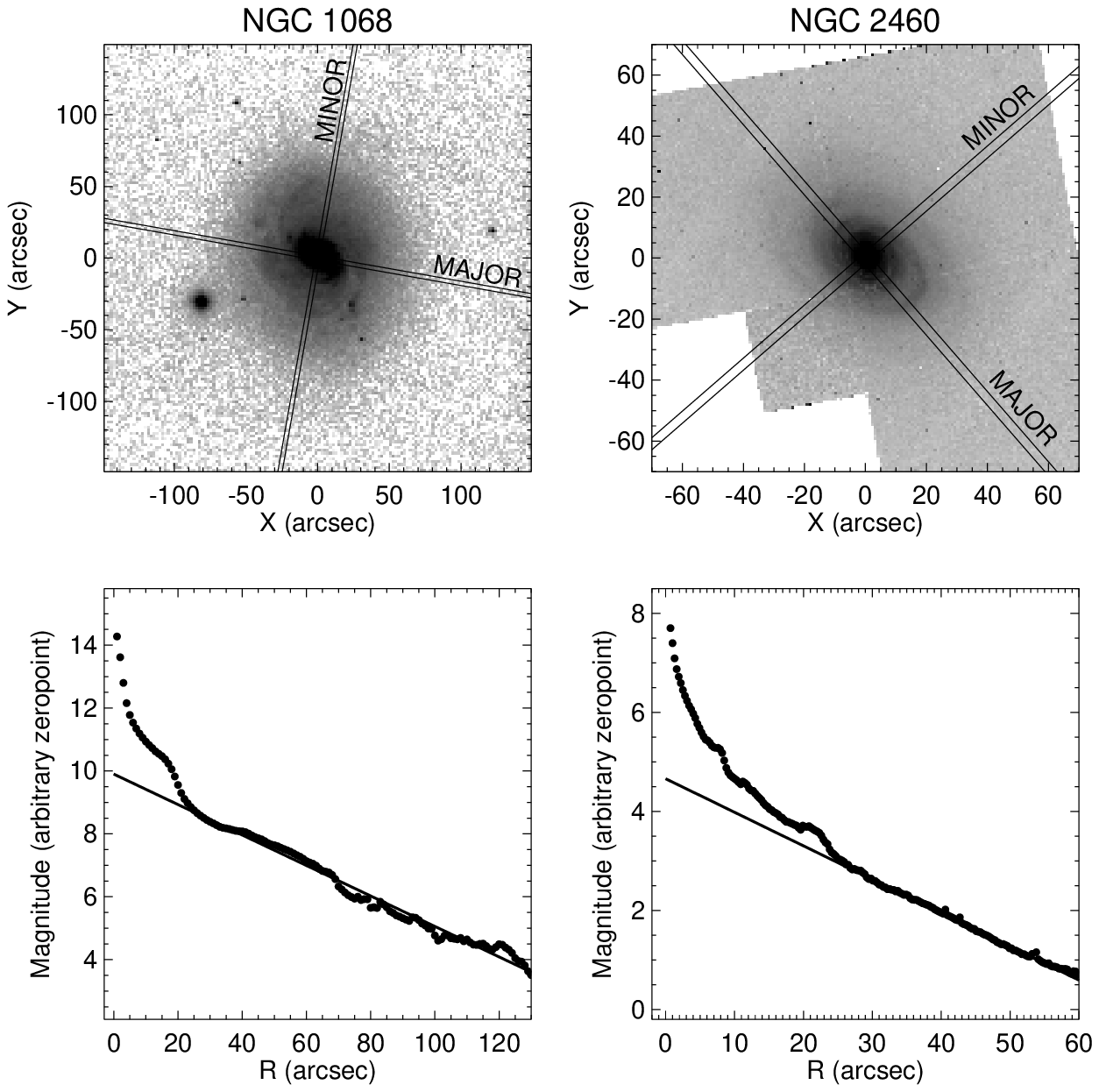}\hfill
\epsfxsize=0.5\hsize
\epsfbox{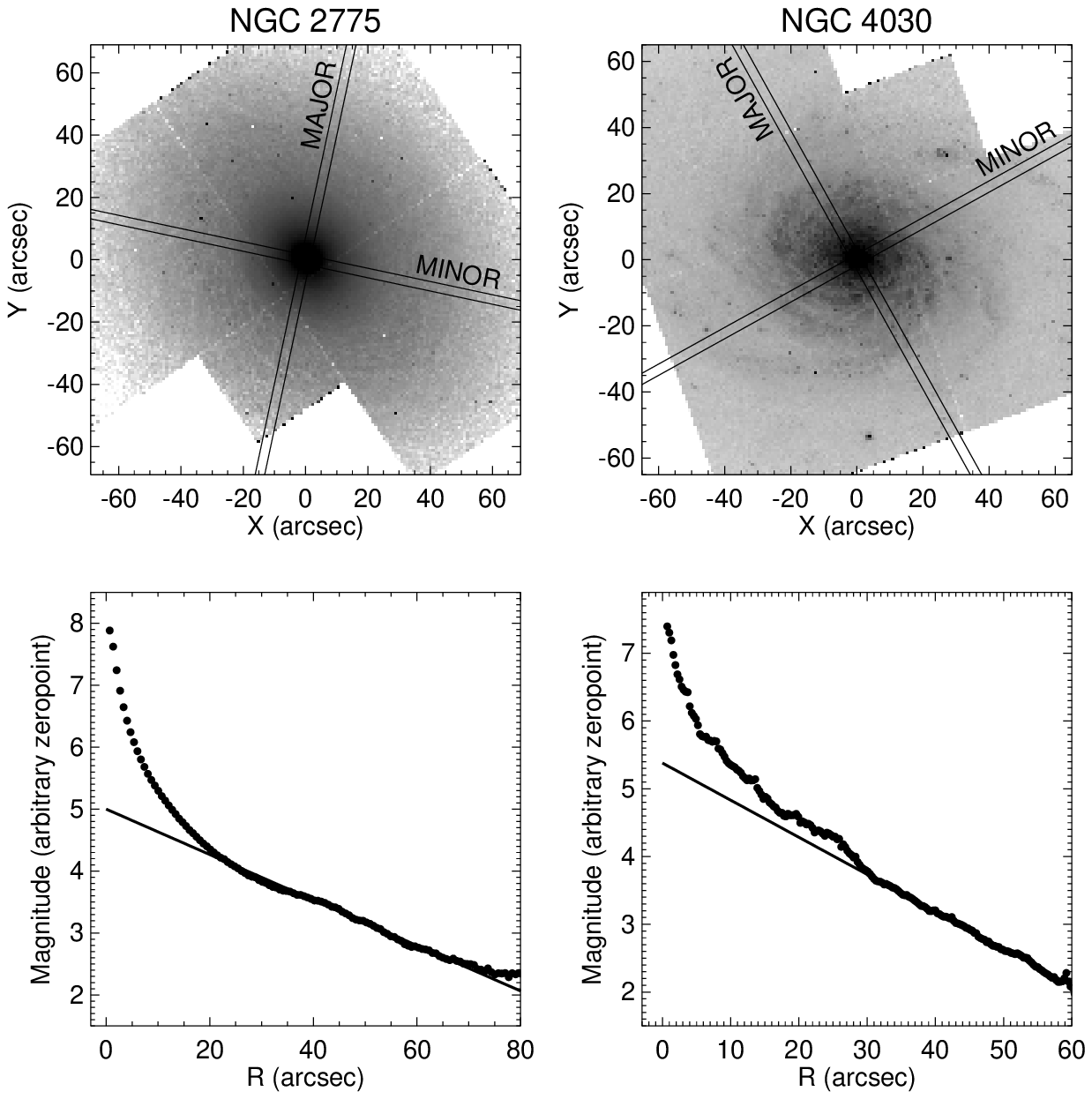}}
\ifsubmode
\vskip3.0truecm
\addtocounter{figure}{1}
\centerline{Figure~\thefigure}
\else\figcaption{\figcapsbdata}\fi
\end{figure}

\clearpage
\begin{figure}
\epsfxsize=0.48\hsize
\hbox to \hsize{\noindent\epsfbox{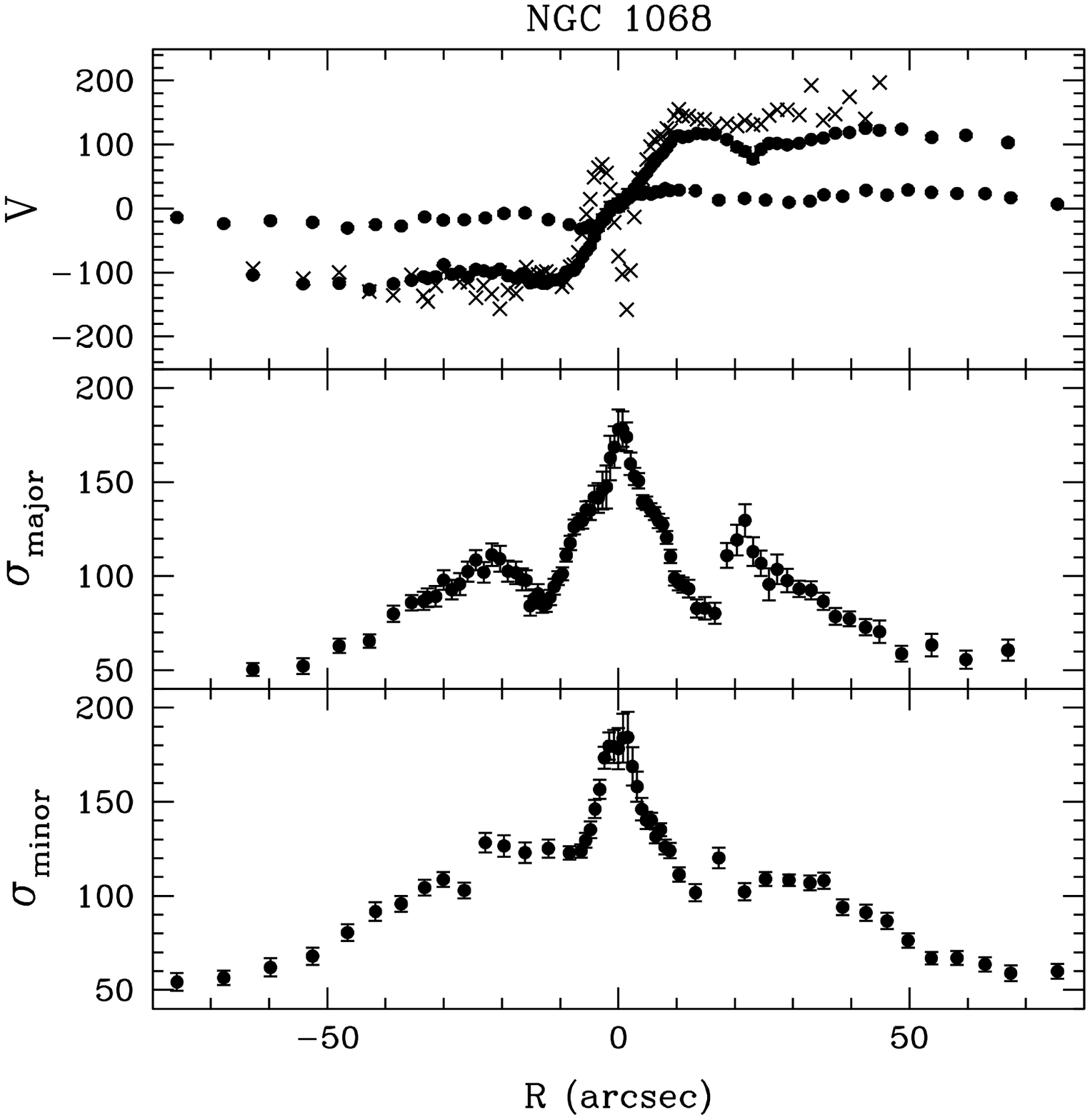}\hfill
\epsfxsize=0.48\hsize
\epsfbox{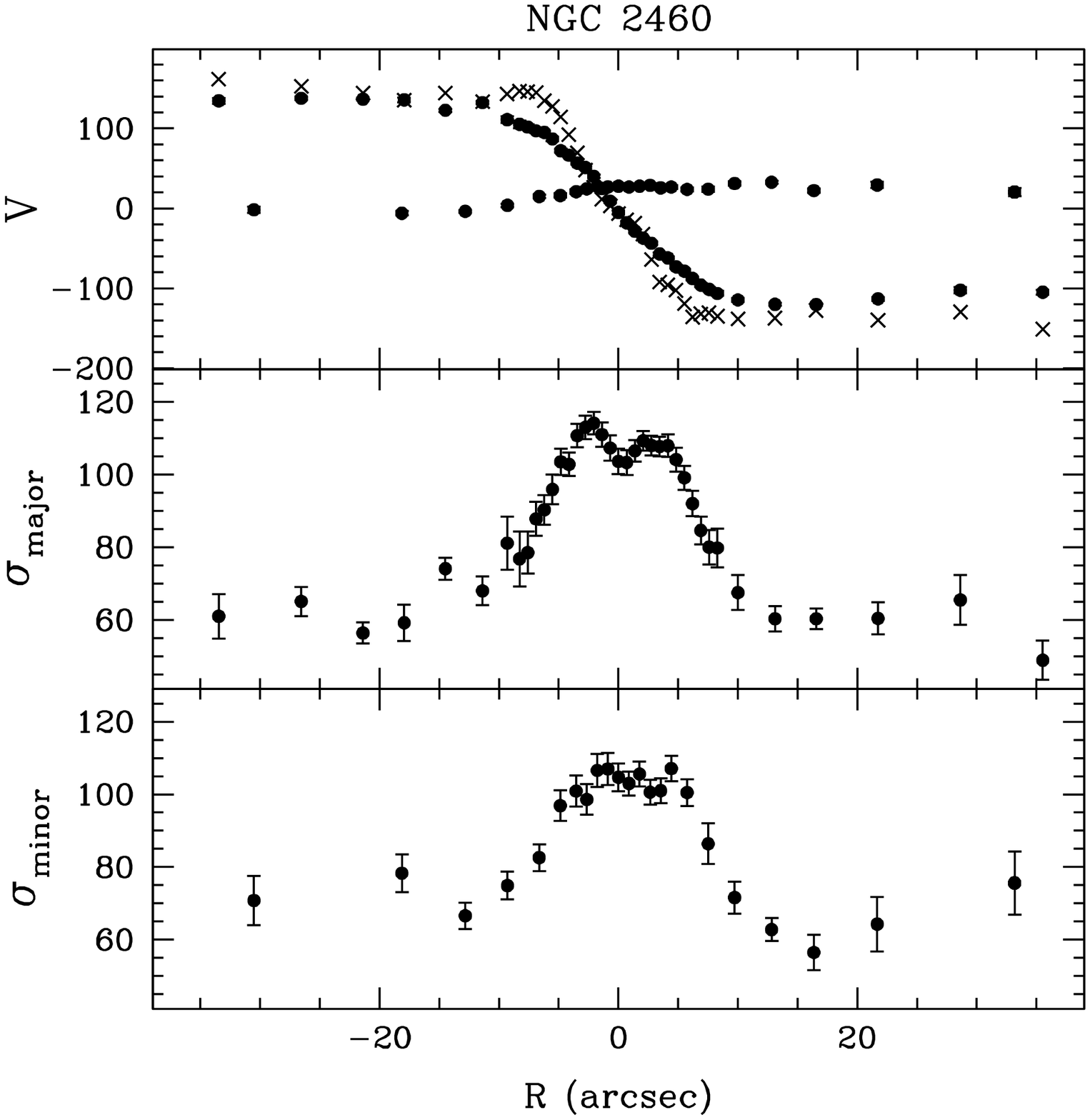}}
\vspace{0.3truecm}
\epsfxsize=0.48\hsize
\hbox to \hsize{\noindent\epsfbox{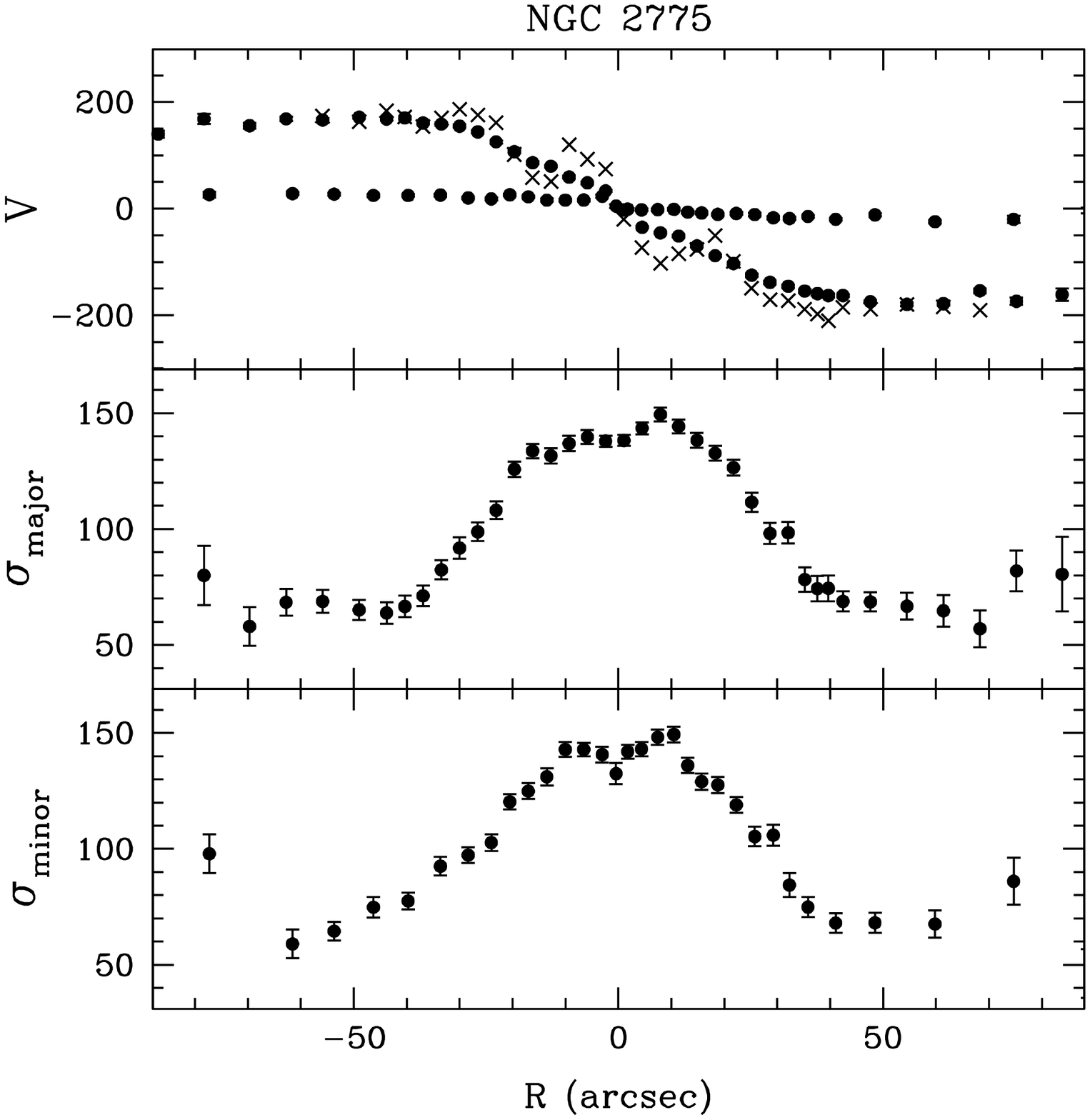}\hfill
\epsfxsize=0.48\hsize
\epsfbox{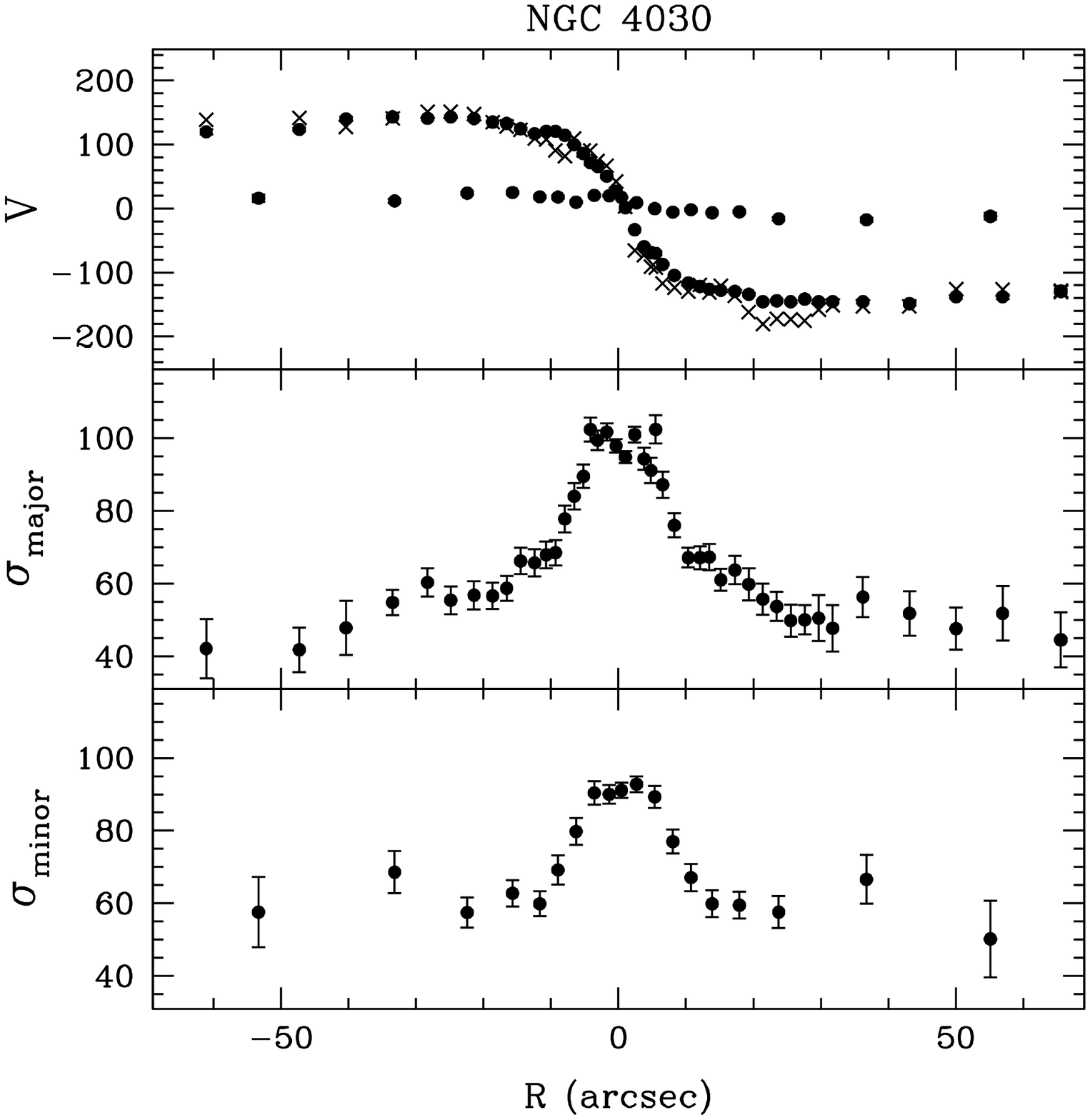}}
\ifsubmode
\vskip3.0truecm
\addtocounter{figure}{1}
\centerline{Figure~\thefigure}
\else\figcaption{\figcapkin}\fi
\end{figure}

\clearpage
\begin{figure}
\epsfxsize=0.48\hsize
\hbox to \hsize{\noindent\epsfbox{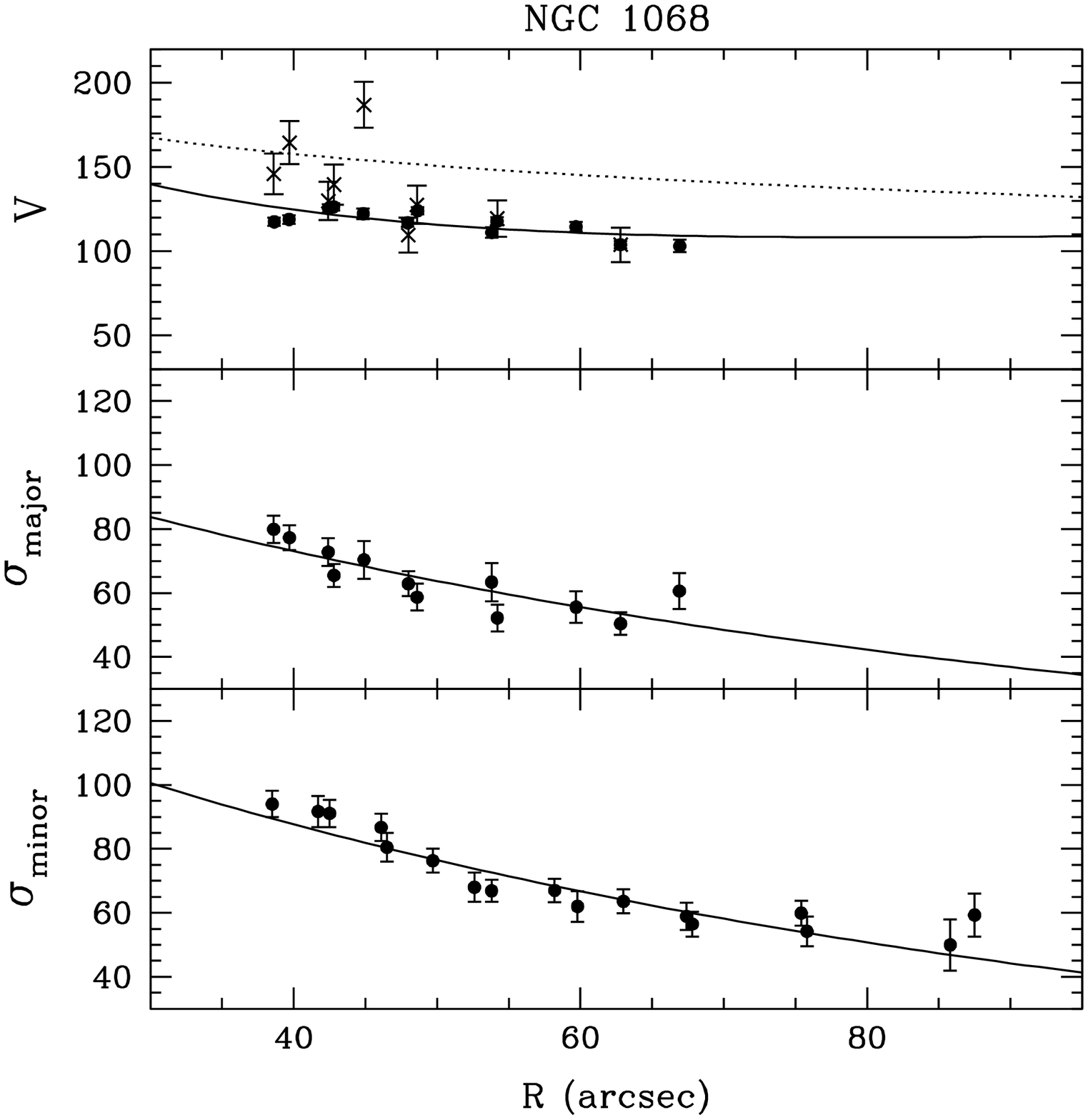}\hfill
\epsfxsize=0.48\hsize
\epsfbox{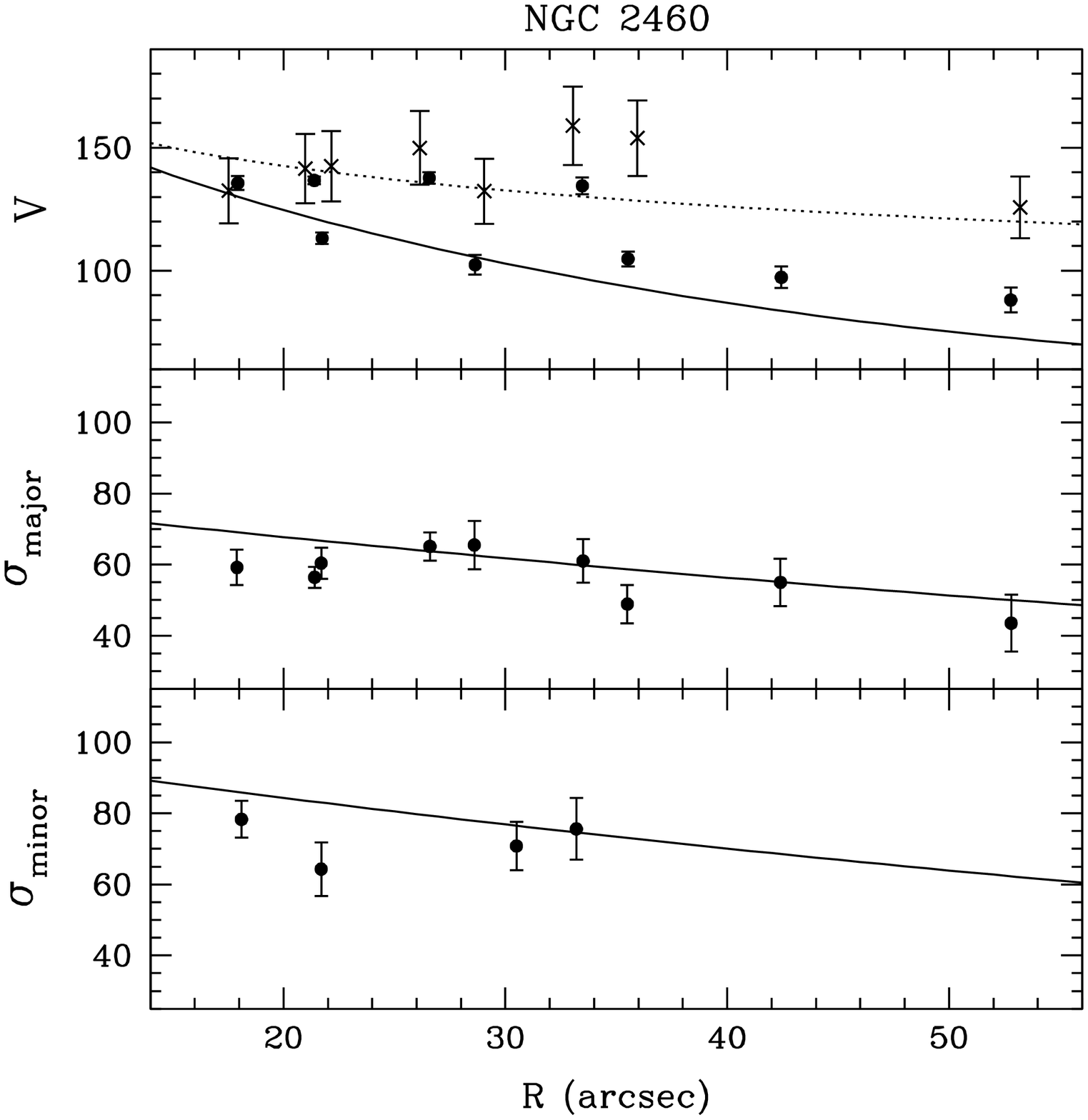}}
\vspace{0.3truecm}
\epsfxsize=0.48\hsize
\hbox to \hsize{\noindent\epsfbox{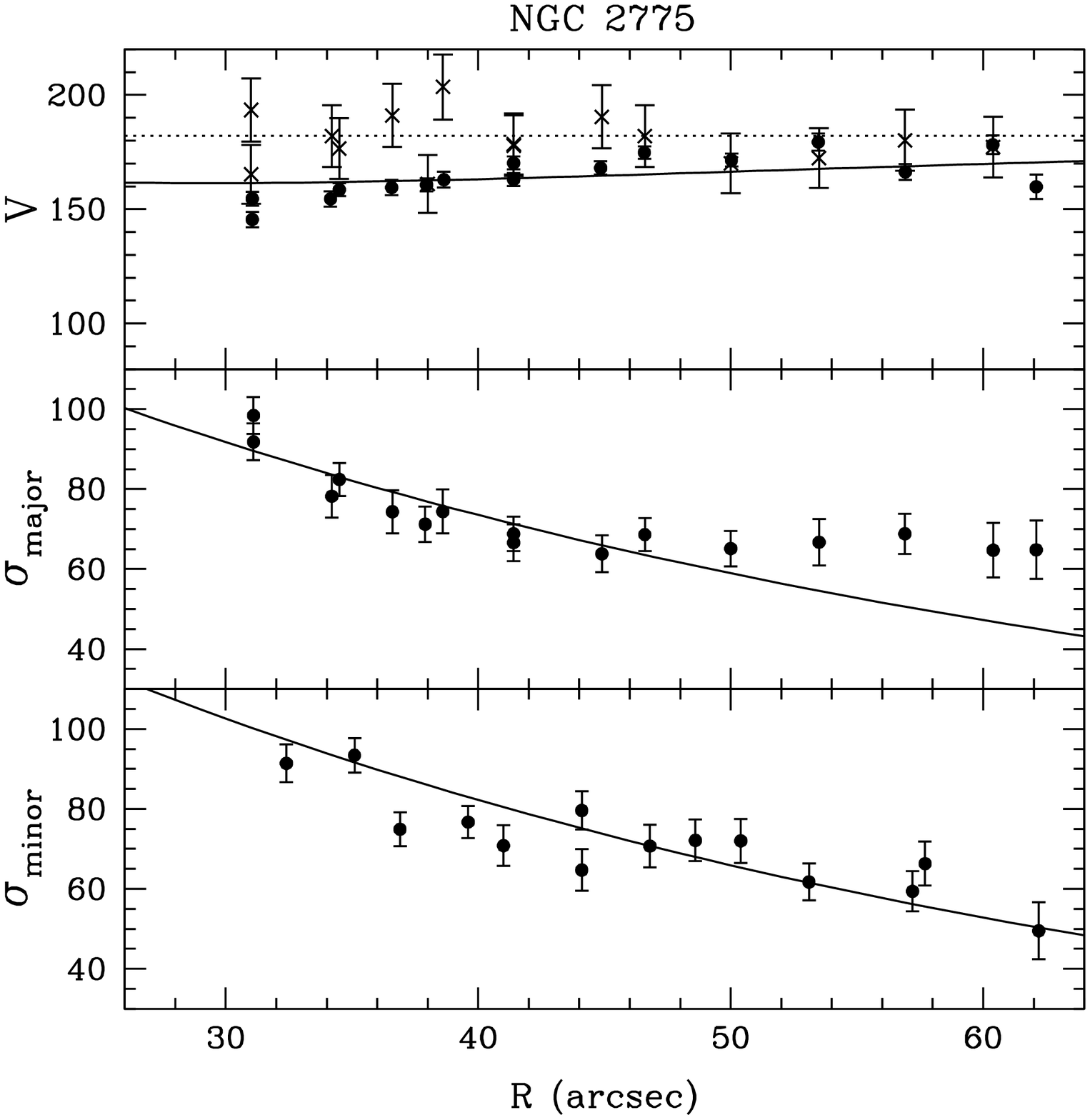}\hfill
\epsfxsize=0.48\hsize
\epsfbox{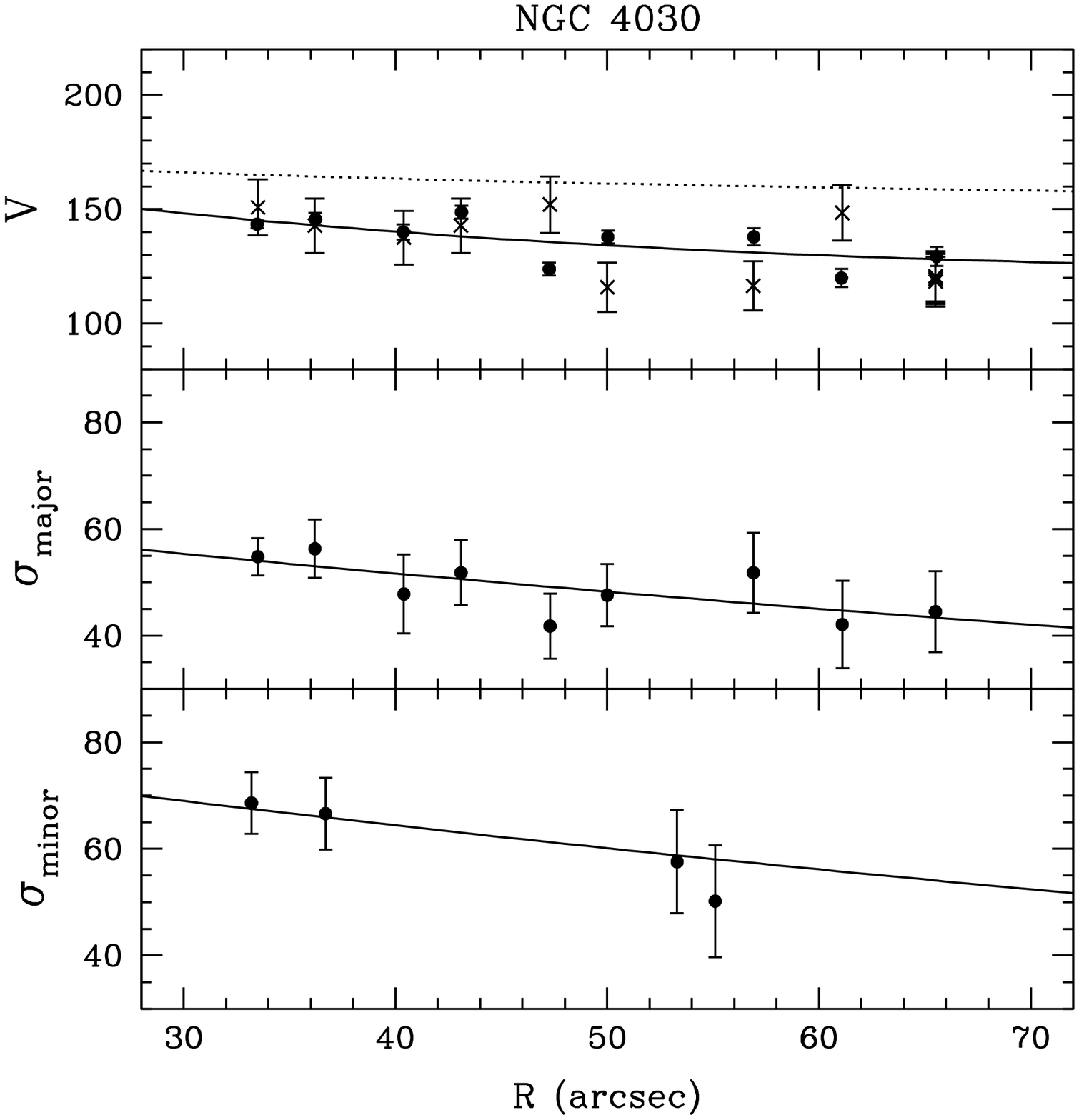}}
\ifsubmode
\vskip3.0truecm
\addtocounter{figure}{1}
\centerline{Figure~\thefigure}
\else\figcaption{\figcapallmodels}\fi
\end{figure}

\clearpage
\begin{figure}
\epsfxsize=15.0truecm
\epsfysize=15.0truecm
\centerline{\epsfbox{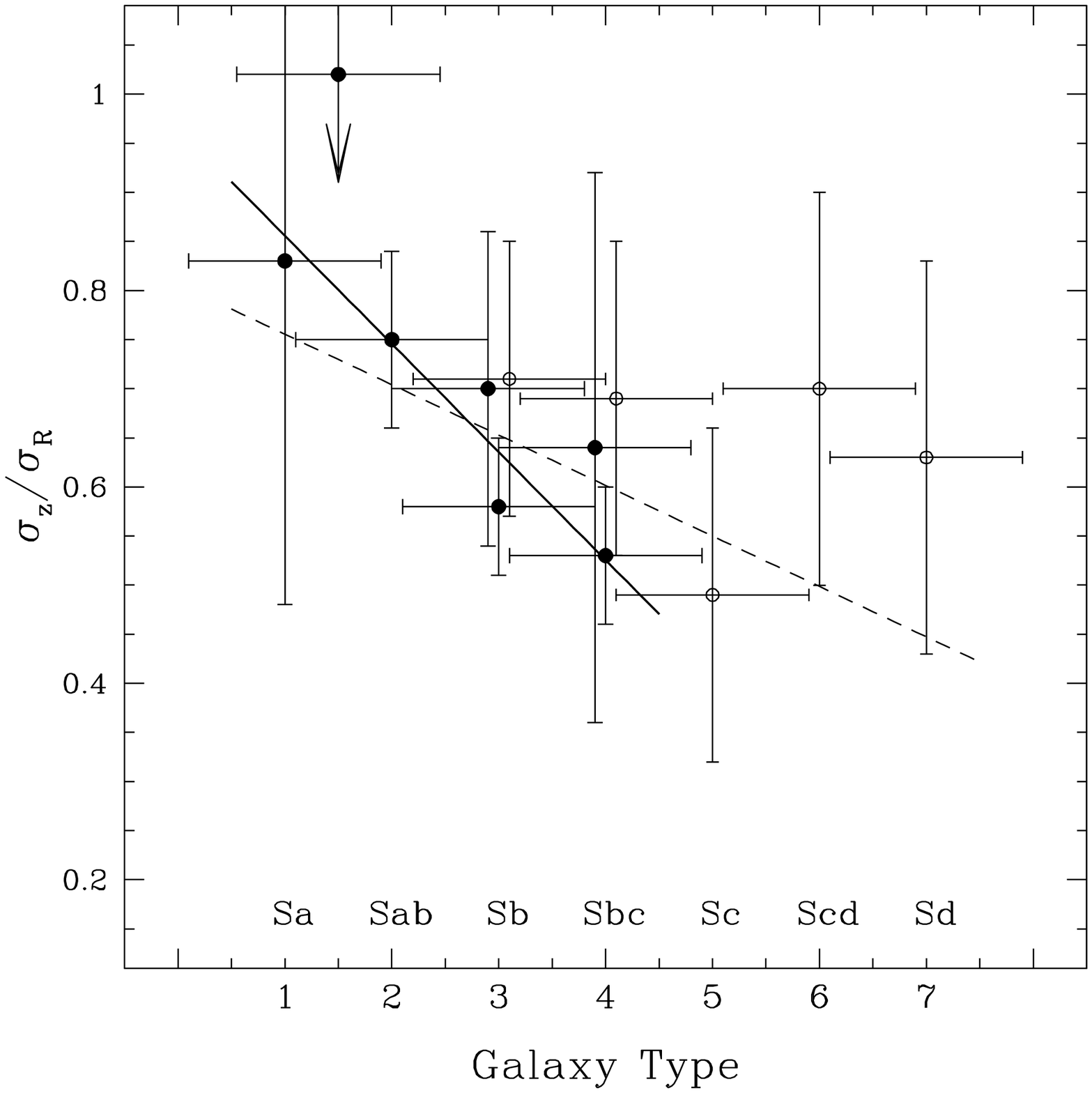}}
\ifsubmode
\vskip3.0truecm
\addtocounter{figure}{1}
\centerline{Figure~\thefigure}
\else\figcaption{\figcapfinal}\fi
\end{figure}



\clearpage
\ifsubmode\pagestyle{empty}\fi

\begin{deluxetable}{lccllc}
\tablewidth{0pc}
\tablecaption{Galaxy Sample}
\tablehead{
\colhead{Galaxy} & \colhead{Type} & \colhead{Redshift} & 
\colhead{Inclination} & \colhead{Scale Length}  & \colhead{Comments} \\ 
\colhead{} & \colhead{} & \colhead{(km s$^{-1}$)} & \colhead{(degrees)} & 
\colhead{(arcsec)}  &  \colhead{}} 
\ifsubmode\renewcommand{\arraystretch}{0.68}\fi
\startdata
NGC 1068   &  Sb      &  1137  &  \ 30 $\pm$ 9$^{\rm 4,5,6,7}$  & \ \ \ 21 $\pm$ 3$^{\rm 7}$  &  $a$ \\
NGC 2460   &  Sa      &  1442  &  \ 46 $\pm$ 7$^{\rm 5,6}$  &  \ \ \ 15 $\pm$ 2$^{\rm 5}$  &  $b$ \\
NGC 2775   &  Sa/Sab  &  1354  &  \ 40 $\pm$ 8$^{\rm 2,5,9}$  & \ \ \ 35 $\pm$ 7$^{\rm 8,9}$  &  $c$ \\
NGC 4030   &  Sbc     &  1460  &  \ 40 $\pm$ 12$^{\rm 3,5,10}$ & \ \ \ 18 $\pm$ 2$^{\rm 1,5}$  &  $b$, $d$ \\
\enddata
\tablecomments{
All redshifts and morphological classifications are from NED.  The
inclinations and scale lengths, unless otherwise noted, are derived
using the process described in Section \ref{s:photo}. Errors on the
inclinations represent the spread in literature values for these
quantities. \\
a) The photometric scale length found here agrees with the value of
21.4 arcsec found by Laurikainen \& Salo (2002) from 2MASS images. \\
b) Other infrared decompositions of this galaxy were not found in the
literature.  Although this scale length can not be compared to other
similar values, it is in keeping with available photographic
decompositions of the galaxy (Grosbol 1985; Baggett et al. 1998). \\
c) The scale length for NGC 2775 quoted here represents the average of
two literature K band decompositions (Moriondo et al. 1998; Mollenhoff
\& Heidt 2001) and that done here.  The error represents the total
spread in these values. \\
d) For NGC 4030, the isophotal ellipse method of finding the
inclination was unsuccessful at arriving at a single value.  The value
quoted here represents the average of available literature values
(Grosbol 1985; Roth 1994; Frei et al. 1996). \\
\\
References: \\
1) Baggett et al. 1998 \\
2) Boroson 1981 \\
3) Frei et al. 1996 \\
4) Garcia-Gomez et al. 2002 \\
5) Grosbol 1985 \\
6) Huchtmeier \& Richter 1989 \\
7) Laurikainen \& Salo 2002 \\
8) Mollenhoff \& Heidt 2001 \\
9) Moriondo et al. 1998 \\
10) Roth 1994
}
\label{t:sample}
\end{deluxetable}

\clearpage
\ifsubmode\pagestyle{empty}\fi

\begin{deluxetable}{lcc}
\tablewidth{0pc}
\tablecaption{Log of Spectroscopic Observations}
\tablehead{
\colhead{Galaxy} & \colhead{Position Angle} & \colhead{Integration Time}\\ 
\colhead{} & \colhead{(degrees)} & \colhead{(minutes)}} 
\ifsubmode\renewcommand{\arraystretch}{0.68}\fi
\startdata
NGC 1068   &  80 - major axis   & 270 \\
           &  170 - minor axis  & 210 \\
 & & \\
NGC 2460   &  41 - major axis   & 240 \\
           &  131 - minor axis  & 300 \\
 & & \\
NGC 2775   &  168 - major axis  & 300 \\
           &  78 - minor axis   & 300 \\
 & & \\
NGC 4030   &  29 - major axis   & 270 \\
           &  119 - minor axis  & 120 \\
\enddata
\tablecomments{Major and minor axis position angles were derived 
from the RC3 catalog or DSS images.}
\label{t:obslog}
\end{deluxetable}

\clearpage
\ifsubmode\pagestyle{empty}\fi

\begin{deluxetable}{ccccccc}
\tablewidth{0pc}
\tablecaption{Best-fit Model Parameters}
\tablehead{
\colhead{Parameter}  & \colhead{NGC 1068} & \colhead{NGC 2460}  &  
\colhead{NGC 2775}  &  \colhead{NGC 4030}  & \colhead{NGC 488}  &  
\colhead{NGC 2985}}  

\ifsubmode\renewcommand{\arraystretch}{0.68}\fi
\startdata
$V_h$ ($\kms$)     & 356 $\pm$ 44             & 218 $\pm$ 24 
 & 283 $\pm$ 4     & 263 $\pm$ 51      & 336 $\pm$ 22      & 249 $\pm$ 16 \\

$\alpha$  		 &  -0.21 $\pm$ 0.03  &  -0.12 $\pm$ 0.15  
 &  set at 0      &  -0.08 $\pm$ 0.06  &  0.21 $\pm$ 0.04  &  0.18 $\pm$ 0.03 \\

$\sigma_{R,0}$ ($\kms$)  &  213 $\pm$ 20      &  110 $\pm$ 12  	   
 &  197 $\pm$ 14  &  105 $\pm$ 23      &  253 $\pm$ 32     &  156 $\pm$ 12 \\

$\sigma_{z,0}$ ($\kms$)  &  124 $\pm$ 9       &  92 $\pm$ 24       
 &  201 $\pm$ 14  &  67 $\pm$ 20       &  164 $\pm$ 27     &  117 $\pm$ 10 \\

$h_{\rm kin}$ (arcsec)   &  72 $\pm$ 6        &  108 $\pm$ 55      
 &  45 $\pm$ 3    &  140 $\pm$ 63      &  38 $\pm$ 4       &  88 $\pm$ 13 \\

\enddata
\tablecomments{
Best-fit parameters derived from the modeling procedure.  The quoted
errors are one-sigma errors.  For NGC~2460, the error includes
systematic errors due to the asymmetric nature of the observed
rotation curve (see Section~\ref{s:2460mod}).  Results for NGC 488 and
NGC 2985 are from improved analyses of the data presented in Gerssen
et al. (1997, 2000).  For NGC~488, we used fit 1 of Gerssen \etal
(1997), which assumes $\sigma_{Rz}$=0, as we do for the galaxies
observed here (see Section~\ref{s:theory}).  For NGC~2775, $\alpha$
was kept fixed at zero.}
\label{t:allresults}
\end{deluxetable}

\clearpage
\ifsubmode\pagestyle{empty}\fi

\begin{deluxetable}{lccc}
\tablewidth{0pc}
\tablecaption{Heating in Sample Galaxies}
\tablehead{
\colhead{Galaxy} & \colhead{Hubble Type} & \colhead{H$_2$ Surface Density} & 
\colhead{$\sigma_z/\sigma_R$} \\ 
\colhead{} & \colhead{} & \colhead{($\Msun \pc^{-2}$)} & \colhead{}} 
\ifsubmode\renewcommand{\arraystretch}{0.68}\fi
\startdata
NGC 2460  &  Sa      &  ...      &    0.83 $\pm$ 0.35 \\
NGC 2775  &  Sa/Sab  &  28       & $<$1.02 $\pm$ 0.11 \\
NGC 2985  &  Sab     &  22       &    0.75 $\pm$ 0.09 \\
NGC 488   &  Sb      &  8        &    0.70 $\pm$ 0.16 \\
NGC 1068  &  Sb      &  39       &    0.58 $\pm$ 0.07 \\
NGC 4030  &  Sbc     &  16       &    0.64 $\pm$ 0.28 \\
Milky Way &  Sbc     &  1.8      &    0.53 $\pm$ 0.07 \\
\enddata
\tablecomments{Velocity ellipsoid ratio data is from 
Table~\ref{t:allresults}, except for the Milky Way value, which is
from Hipparcos data of the solar neighborhood (Dehnen \& Binney 1998).
Molecular content information is estimated from CO measurements in
Young et al. (1995).  No CO measurements are available for
NGC~2460. For the solar neighborhood, the value of the H$_2$ surface
density is from Clemens et al. (1988).}
\label{t:gas}
\end{deluxetable}


\end{document}